Accepted Version
Publication date: May 2023

Embargo: No Embargo (Accepted Version, under conditions), 24 Months (Published Version)

European Union, Horizon 2020, Grant Agreement number: 857470 — NOMATEN — H2020-WIDESPREAD-2018-2020

DOI: https://doi.org/10.1016/j.nimb.2023.02.036

# Surface and in-depth structural changes in nuclear graphite irradiated with noble gases described with Raman imaging

Magdalena Gawęda[1],*, Magdalena Wilczopolska[1], Kinga Suchorab[1], Małgorzata Frelek-Kozak[1], Łukasz Kurpaska[1], Jacek Jagielski[2,3]

*[1] NOMATEN CoE, NOMATEN MAB, National Centre for Nuclear Research, 7 Andrzeja Sołtana Street, 05-400 Świerk-Otwock, Poland*

*[2] National Centre for Nuclear Research, 7 Andrzeja Sołtana Street, 05-400 Świerk-Otwock, Poland*

*[3] Łukasiewicz Institute for Microelectronics & Photonics, Al. Lotników 32/46, 02-668 Warsaw, Poland*

**corresponding author, email: Magdalena.Gaweda@ncbj.gov.pl*

4th Generation high-temperature gas-cooled nuclear reactors (HTGR) are regarded as possible sources of industrial heat in Poland and Europe, allowing for a substantial reduction of the dependency on gas and coal import. It is mainly due to their safety of use, reliability and economy in a current energetic crisis. In this work, graphite, as a primary construction material and neutron moderator in HTGR, was evaluated before and after ion irradiation since its properties depend on the material's structure and purity. Commercial graphite materials (IG-110, NBG-17) and the laboratory's in-home material were chosen for the exemplary samples. The structural damage in HTGR was simulated with energetic $Ar^+$ and $He^+$ ions with fluencies from 1E12 to 2E17 ion/cm$^2$. Raman imaging was chosen to assess radiation damage build-up: the crystallites' evolution, occurrence and types of defects. The recorded evolution showed stronger disordering of the material with heavier $Ar^+$ ions than with $He^+$.

*Keywords: graphite, HTGR, irradiation, Raman imaging, KMC*

## Introduction

Graphite is the essential construction material for 4th Generation high-temperature gas-cooled reactors (HTGR). It is used because of its beneficial parameters: neutron moderation, high-temperature stability, thermal conductivity, and passively safe fuel reactivity control [1]. Without doubt, the main parameters defining the ability of graphite for the use in a nuclear reactor are its strength, purity, crystallinity and level of structural disorder. However, it is not the case that graphite with the highest purity and large crystallites of high order will have the best functional properties, especially mechanical due to the material anisotropic character [2]. Nevertheless, there is a need for the method of evaluation of the material's structure and its reaction on radiation damage. Here Raman spectroscopy comes to hand. The method is well known and widely used in materials science, especially due to its non-destructive, non-contacting character and fast measurement. It is possible to identify particular compounds from an unique set of bands corresponding to energies of particular bonds in the compound – the so-called “fingerprint”. Moreover, the method can detect subtle structural changes or residual stress [3]. Nowadays, the Raman spectrometers are merged with microscopes [4, 5, 6]. Such a constructions provides control of the measurement spot and (with the control of the microscope stage) builds the network of Raman spectra of the chosen pattern – Raman imaging. In that way, it is possible also to observe structural differences between sections of the material and the character of the interphases. Moreover, many microspectrometers are confocal and allow in-depth measurements without special sample preparation [7].

The Raman spectrum of graphite consists of a number of characteristic bands (Fig. 1), among which the so-called G and D bands are the most pronounced. The first mentioned G band positioned at approximately 1580 $cm^{-1}$ corresponds to the carbon-carbon bond in $sp^2$ hybridization in graphite sheets and is associated with the material's order. At the same time, the D band at approximately 1350 $cm^{-1}$ relates to the $sp^3$ C-C bond with a dangling bond which might be found on the edges of the layers and neighboring defects [1]. Therefore, the D band is connected with the disorder allowing for the evaluation of the disorder from the ratio between intensities of these two bands ($^{I_D}/_{I_G}$ ratio) [8, 9, 10]. Not only the level but also the character of the defects might be detected [11].

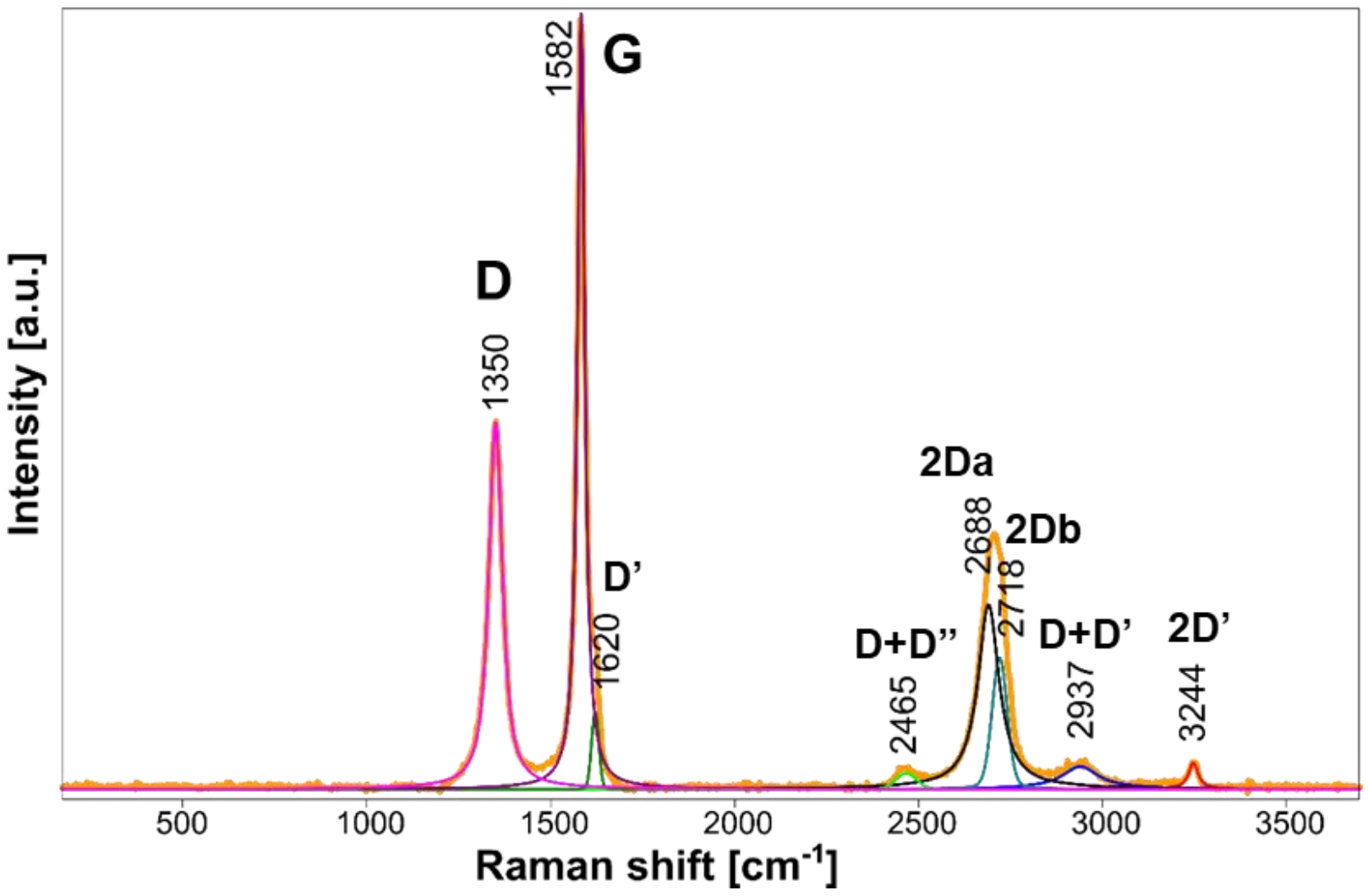

*Figure 1. Exemplary Raman spectra of graphite with fitted component bands [1]*

However, in terms of the proposed application, the exact level of the material's initial structure is not a key factor. It is its evolution when exposed to harsh work conditions in the reactor that matters. To evaluate material suitability, it is necessary to describe the structural evolution of the material in three aspects: the influence of irradiation, high temperature, and a combination of these factors acting simultaneously. Some studies on the response of carbon-based materials to irradiation with various factors might be found in the literature. As an example, the amorphisation of highly-oriented pyrolytic graphite under irradiation with $D^+$ and $He^+$ in temperatures up to 700 °C was investigated, indicating the temperature-dependent threshold value of fluence causing critical structural damage [12]. The step-like order dependence on fluence was also demonstrated by irradiation with a broad range of ions ($^{7}Li$, $^{9}Be$, $^{11}B$, $^{12}C$, $^{31}P$, $^{75}As$), along with the temperature as the counterfactor to the amorphization [13]. Moreover, the evolution of physicochemical and functional parameters was previously analysed. For example, the rise of Young's modulus and hardness values with high-energetic protons irradiation was shown to be strictly connected with a significant reduction in crystal size [14].

In this work some steps were made to characterize graphite homogeneity and describe the behavior and structural stability of the material after irradiation with two types of ions at elevated temperatures. The WITec alpha 300R Raman spectrometer was used for a detailed examination of the structure of commercially available graphites (IG-110 and NBG-17) along with the laboratory's in-home material. Results of this research will help to elucidate graphite evolution upon irradiation, describe its homogeneity and help prepare a methodology of quick and efficient evaluation of the carbon-based materials for the use in the next-generation nuclear reactors.

## Methodology

Two commercially available graphites were chosen: IG-110 (Toyo Tanso, Japan) and NBG-17 (SGL Carbon, Germany). They were compared with the in-home NCBJ material, the same type that one used previously for the EWA reactor construction. It was classified as ASR-0RB grade, mainly based on mechanical properties [15]. Before further proceedings, samples were polished with SiC sandpaper (gradation 4000) and washed in an ultrasonic washer in ethyl alcohol to remove residues of abrasives.

Studied materials were irradiated with 150 keV argon and helium ions at 400 °C up to max. fluence of 2E17 ions/cm$^2$. The use of $Ar^+$ ions aims to develop a high dpa level and simulate the elastic collisions. In contrast, irradiation with $He^+$ is aimed to simulate more inelastic interactions [16]. Detailed information on the irradiation, including estimated displacements per atom (dpa) and depth of dpa peak [μm] estimated with the SRIM software [16], are presented in Table 1.

*Tab. 1. Ion irradiation parameters of graphitic materials*

| **150 keV $Ar^+$ and $He^+$ at 400 °C** | | | | |
|---|---|---|---|---|
| **Fluence [ion/cm$^2$]** | **Estimated displacements per atom (dpa)** | | **Estimated depth of dpa peak [μm]** | |
| | Ar | He | Ar | He |
| 0 | - | - | - | - |
| 1E12 | 0,0129 | 0,00005 | 0,081 | 0,63 |
| 1E16 | 12,916 | 0,50519 | | |
| 2E17 | 258,32 | 10,1038 | | |

Raman imaging was performed using confocal WITec alpha 300R spectrometer (Oxford Instruments) controlled by the WITec Control 5.2 software and equipped with a 532 nm laser. Maps had dimensions as follows: for surface 25 × 25 μm in X and Y directions (50 points per 50 lines, point every 0.5 μm) and for in-depth 25 × 25 μm in X and Z directions (50 points per 50 lines, point every 0.5 μm). They were acquired with Zeiss LD EC Epiplan-Neofluar Dic 50x/0.55 lense, 600 lines/mm grating and 3 s acquisition time. Laser power was set to 10 mW (surface) and 20 mW (in-depth). Series of consecutive measurements in the same spot confirmed, that this power do not lead to excessive heating of the sample, hence to annealing effects induced by measurement. Post-measurement treatment of the data was done in the WITec Project FIVE 5.2 software. It included cutting spectral region of interest, baseline correction, cosmic ray removal (CCR), band fitting with PseudoVoigt function (included in the WITec software), basic calculations, and chemometric analysis (K-means clustering, KMC [17]).

As mentioned above, calculations covered estimating $I_D/I_G$ ratio, where $I_D$ represents the intensity of the D band and $I_G$ stands for the intensity of the G band. In this work, the maximum intensities of bands were used in calculations. The obtained results allowed us to follow degree of disorder in the examined materials as well as the character of the damage induced by irradiation.

The size of crystallites was also analyzed as another indicator of rising disorder in graphite and the material's amorphization. On Raman spectra, this effect is indicated by band's full width at half maximum (FWHM). Calculations for pristine and irradiated samples were peformed accordingly to Eq. 1 [18]. The parameter related to phonon dispersion and decay length values of 195 and 32 nm were taken, respectively from [18].

$$L_a = \frac{l_C}{2} \ln\left[\frac{C}{\Gamma_G^A(L_a) - \Gamma_G^A(\infty)}\right] \qquad \text{(Eq. 1)}$$

| | | |
|---|---|---|
| where: | $C$ | is a parameter related to the phonon dispersion $\omega(q)$, |
| | $l_C$ | is the full decay length [nm], |
| | $\Gamma_G^A(L_a)$ | is FWHM of the G band of the given spectra, |
| | $\Gamma_G^A(\infty)$ | is the minimum FWHM for a pristine sample. |

## Results and discussion

Interpretation of the Raman imaging results might be challenging due to the vast amount of data collected in a single measurement. The maps presented in this article contain 50 points per 50 lines what results in 2500 single spectra per image. The critical point during data evaluation is the simultaneous analysis of all spectra from one map. It has been done using a dedicated WITec Project FIVE 5.2 software. The first step after processing was to fit D and G bands and, based on the obtained intensity values, calculate $I_D/I_G$ ratio. Fig. 2 presents the results of Raman imaging of the surface and in-depth measurements obtained for the pristine and Ar-irradiated samples. The used color scale intuitively shows fluctuating ratio values: the brighter the color – the higher is the ratio and level of disorder.

Analysis of the distribution of the ${}^{I_D}/_{I_G}$ ratio give a picture of the non-uniform structure of the material. The lowest values of the ratio characterize pristine materials. NBG-17 graphite has the highest level of order. On the surface of the NCBJ graphite highest values of the ratio were found, in a cluster-like regions. For the lowest fluence used, a rise of the ratio is significant only in the case of NBG-17. Higher fluencies cause significant disordering of the material. However, NBG-17 seems less affected at the 1E16 ion/cm$^2$ fluence than other materials. Higher disorder results in the shape of the spectra being more complicated and the built-in software is unable to fit bands properly. That causes errors in the performed calculations. Lack of fitting results in singular black spots. That is particularly visible at the bottom of the maps of in-depth profiles.

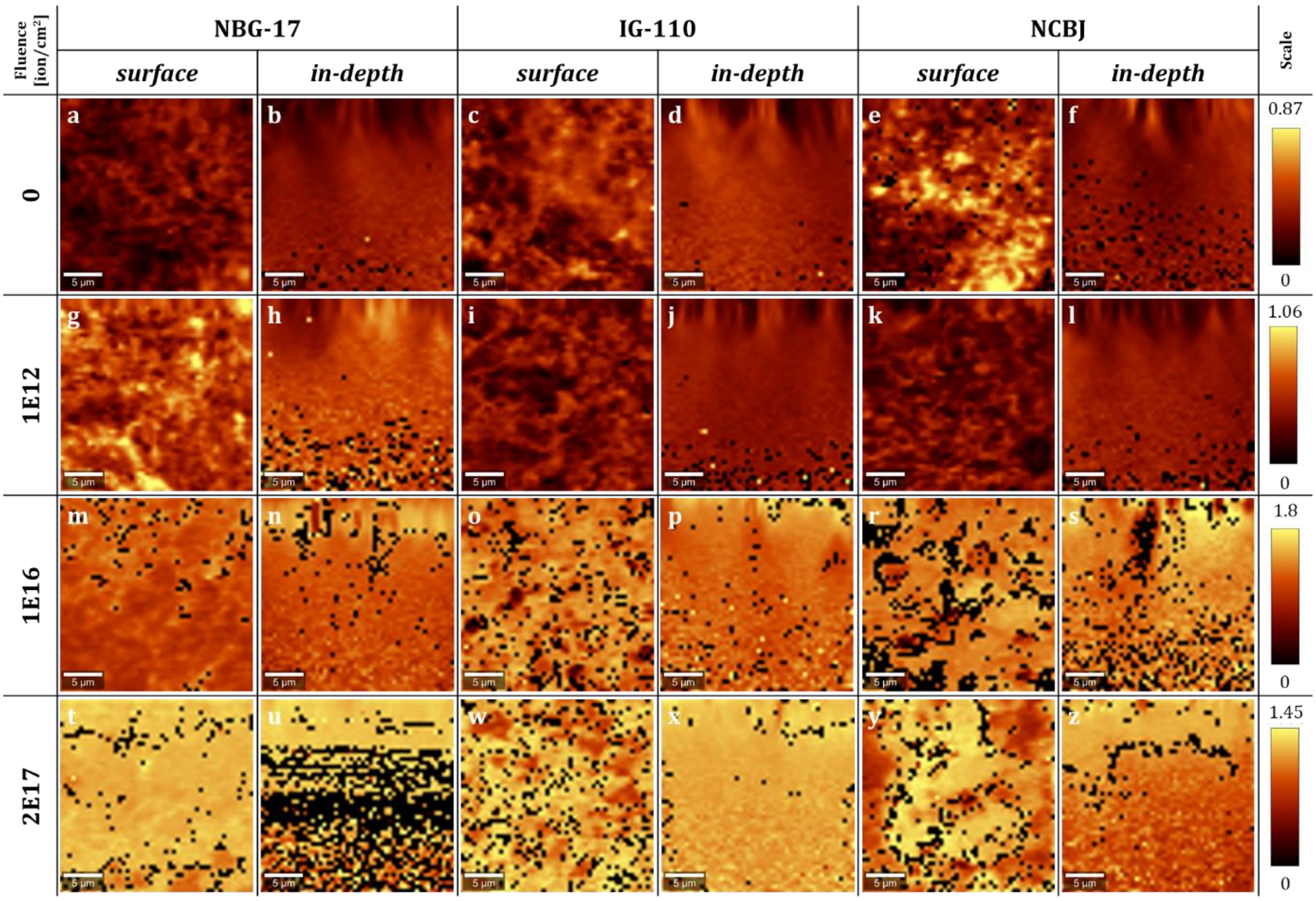

*Figure 2. Distribution maps of the ${}^{I_D}/_{I_G}$ ratio calculated based on Raman imaging of pristine samples and graphite irradiated with $Ar^+$ ions*

After irradiation with helium ions (Fig. 3), the structure of the materials is also distorted by ions. However, since argon ions are much heavier than helium, the significant disorder of the structure for $He^+$ appears at much higher ion fluencies. Here also, NBG-17 seems more affected by irradiation than IG-110 and NCBJ graphite at low fluence. At maximum fluence, the results obtained for all samples become comparable.

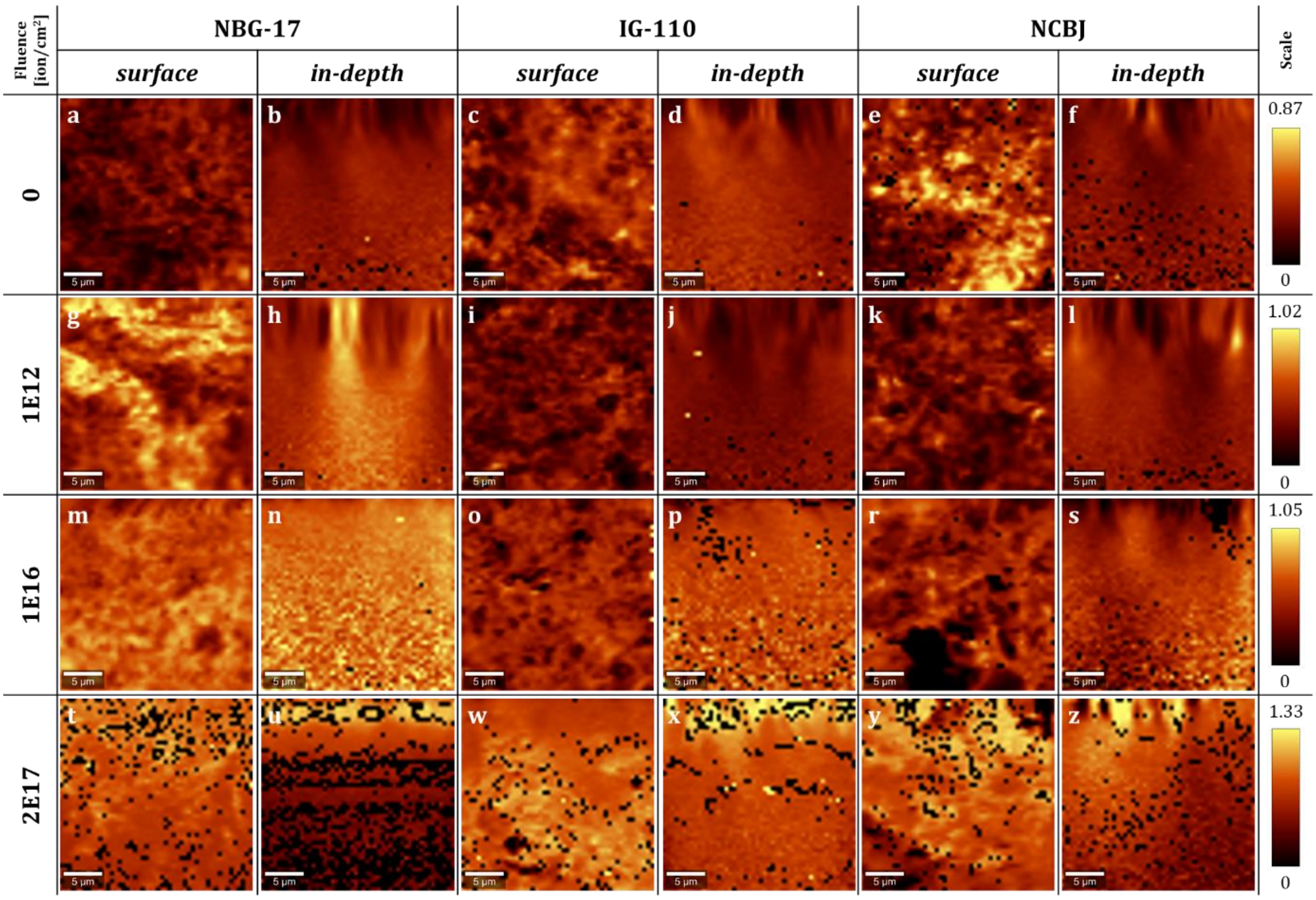

*Figure 3. Distribution maps of the $^{I_D}/_{I_G}$ ratio calculated based on Raman imaging of pristine samples and graphite irradiated with $He^+$ ions*

Fig. 2 and 3, along with the surface measurements, present an in-depth analysis, possible due to the use of the confocal spectrometer. For a better quality of the results, the higher laser power was used. However, quality of the spectra is significantly reduced in the bottom parts of the map. Interpretation and comparison with the surface results become thus more difficult. That is only one of many reasons to use chemometric analysis for further investigation. Fig. 4 presents KMC maps of the NBG-17 samples (pristine, irradiated with $Ar^+$ and $He^+$ up to 1E16 ion/cm$^2$ ) with extracted Raman spectra of particularly distinguished clusters and spectra averaged from all points. The colors were chosen to indicate relative (for a single map) level of disorder based on the $^{I_D}/_{I_G}$ ratio. Blue indicates areas of the highest order (the lowest ratio), orange – is the highest disorder (the highest ratio), and green represents areas of the intermediate character. There are also other colors used for the regions of more distinguishable features: red for impurities, dark and bright pink for areas of higher crystallinity than the rest of the irradiated samples, and azure to highlight regions of well-defined narrow bands indicating crystallinity, yet with the high intensity of D band comparing to G.

Presented in Figure 4, KMC maps for NBG-17 graphite show inhomogeneity of the materials on the surface and in-depth. We can observe significant differences between spectra of samples irradiated with argon and helium ions. After $Ar^+$ irradiation, bands on the spectra are shifted and broadened, which indicates structural changes towards amorphization, including reduction of the crystalline volumes. After the same fluence, but of $He^+$ irradiation, relative intensities of D and G bands are changed, yet bands remain narrow. That suggests a lack of changes in the crystallite size. This aspect will be considered in the following part of this paper.

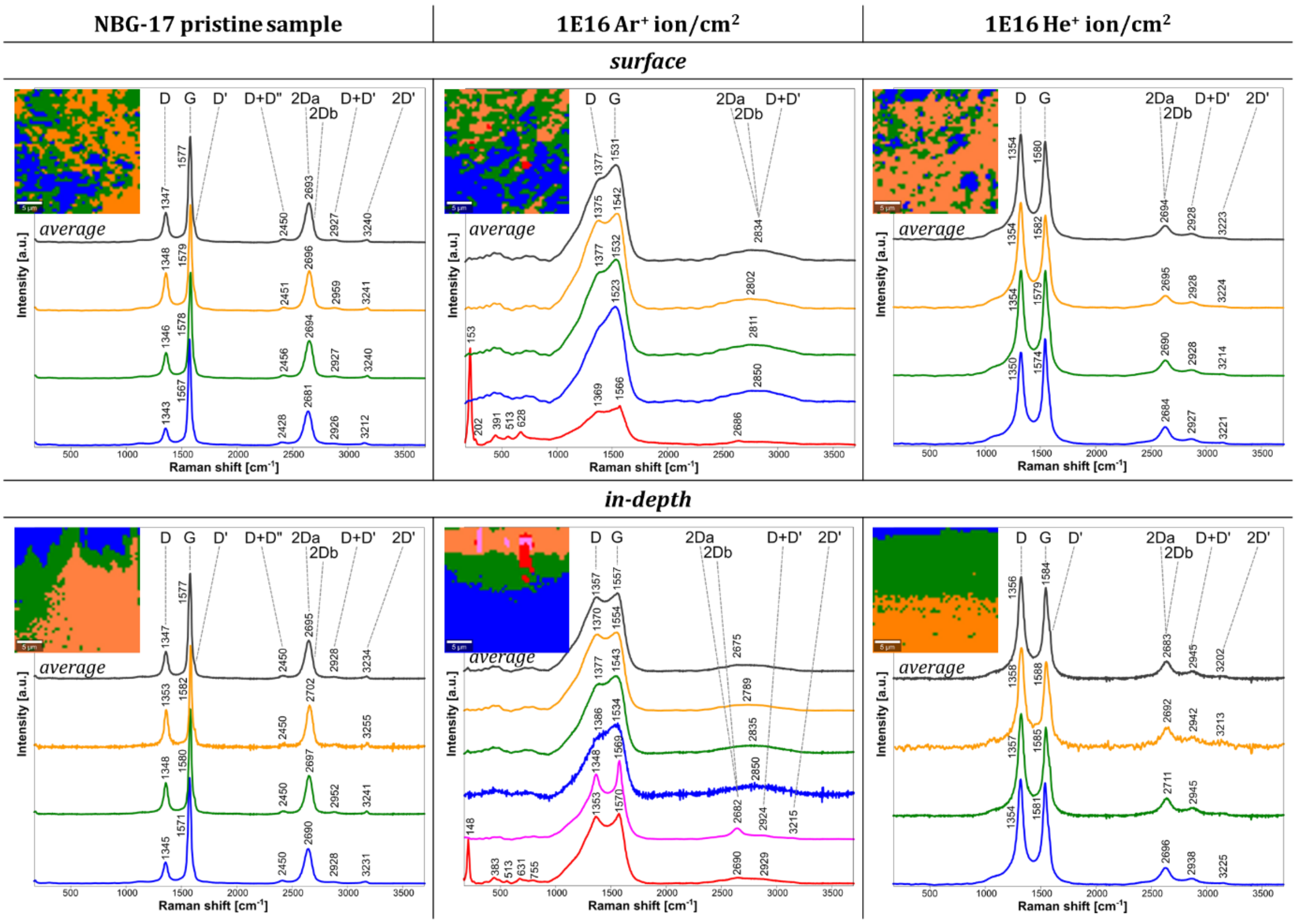

*Figure 4. Exemples of KMC analysis: maps with extracted spectra of particular regions, followed by average spectra*

Results of the KMC analysis for all investigated samples are summarised in Fig. 5 and 6, respectively, for $Ar^+$ and $He^+$ ions compiled with the pristine samples. All of them exhibit inhomogeneity.

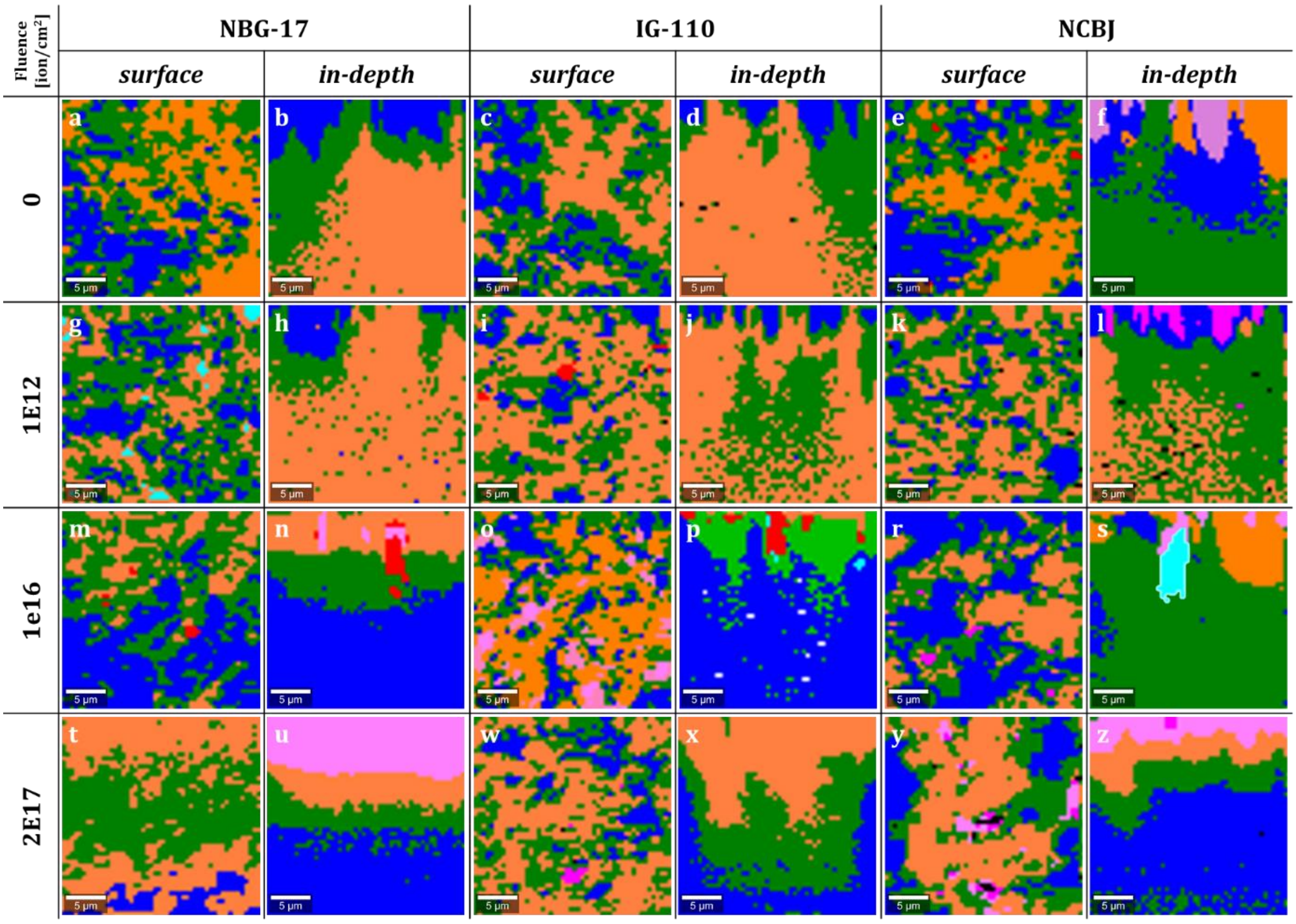

Figure 5. KMC analysis of pristine and Ar-irradiated samples

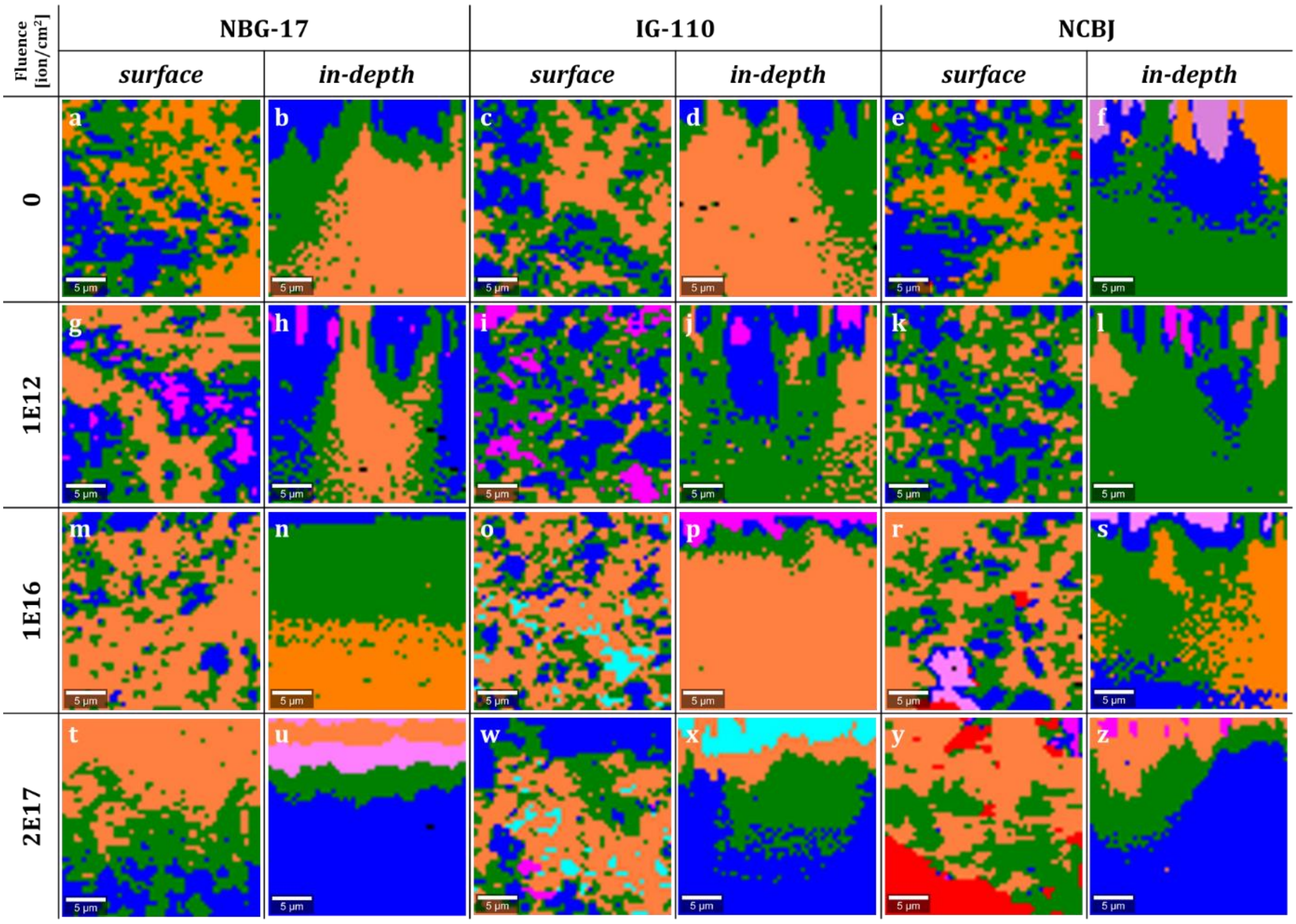

Figure 6. KMC analysis of pristine and He-irradiated samples

More detailed analysis of the $^{I_D}/_{I_G}$ ratio is presented and summarised in Fig. 7. Diagrams present values for particular regions identified during KMS analysis, along with one calculated for average spectra marked black. Pristine materials are similar. Differences are visible already at low fluence, since the range of the ratio's values is broadened and gets wider with the irradiation fluence. However, the average values for all the materials are very similar. Values of the $^{I_D}/_{I_G}$ ratio and observation of Raman spectra profile for $Ar^+$ and $He^+$ irradiated samples suggest the very similar character of defects in all samples: vacancies and edge dislocations [11]. However, in NCBJ and IG-110 irradiated with $He^+$ vacancies are less pronounced [11] [19]. A more detailed description of defects and evolution measurements of the samples with intermediate fluence values should be the focal point of future studies.

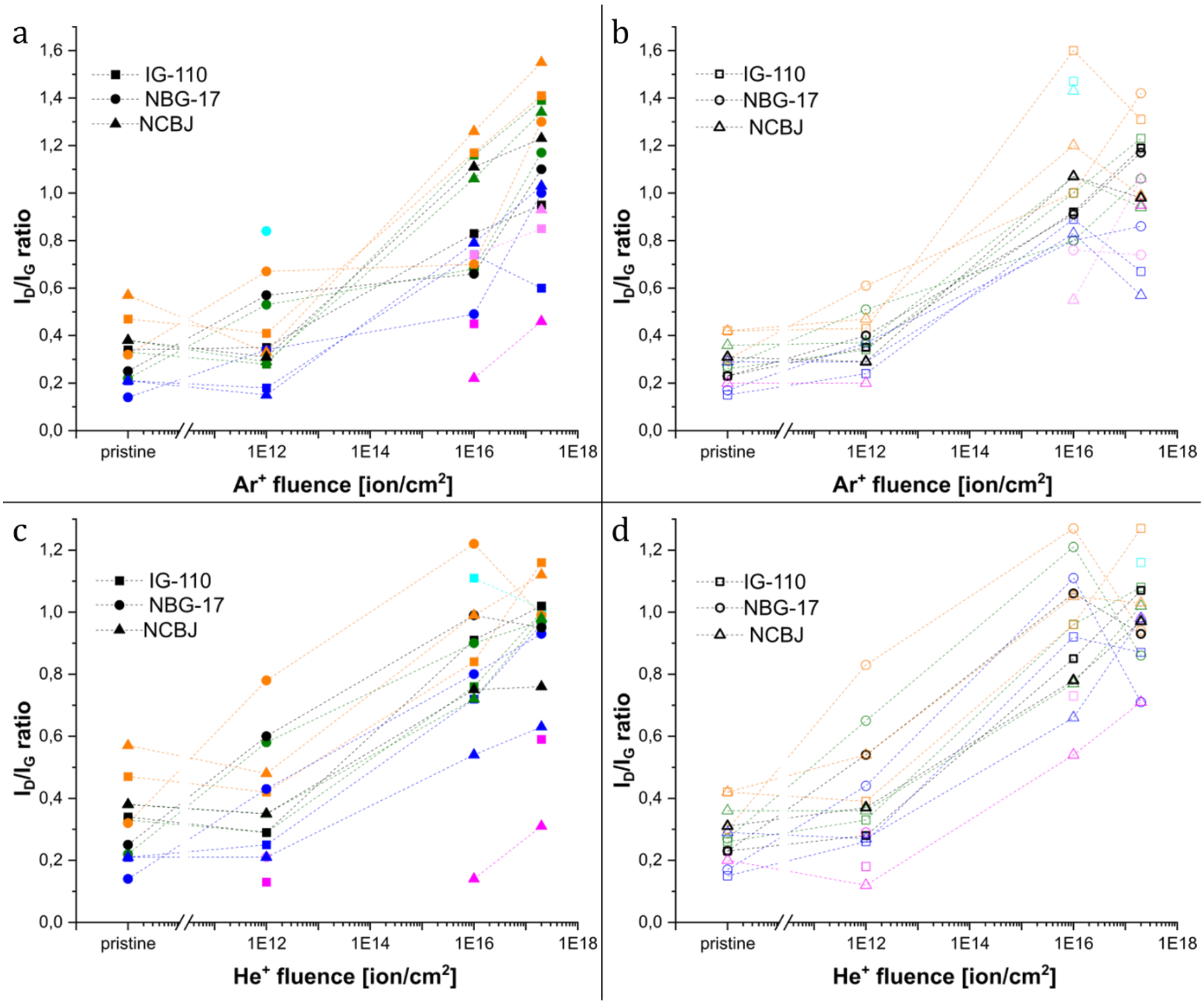

*Figure 7. The $^{I_D}/_{I_G}$ ratios for particular regions (blue, green, orange, pink, light pink and azure) with values for average spectra (black); dashed lines are provided for better eye guidance*

The size of crystallites was calculated as another indicator of raising disorder in graphite. Diagrams present estimated crystallite diameters for surface and in-depth measurements of pristine samples and irradiated with $Ar^+$ and $He^+$ ions (Fig. 8). This result shows the best evidence of the occurring tendencies in evolution. The medium fluence of heavy $Ar^+$ ions already caused a reduction of crystallites to the minimum value of 10 – 15 nm. In the case of light $He^+$ ions damage at 1E16 ion/cm$^2$ is not yet critical.

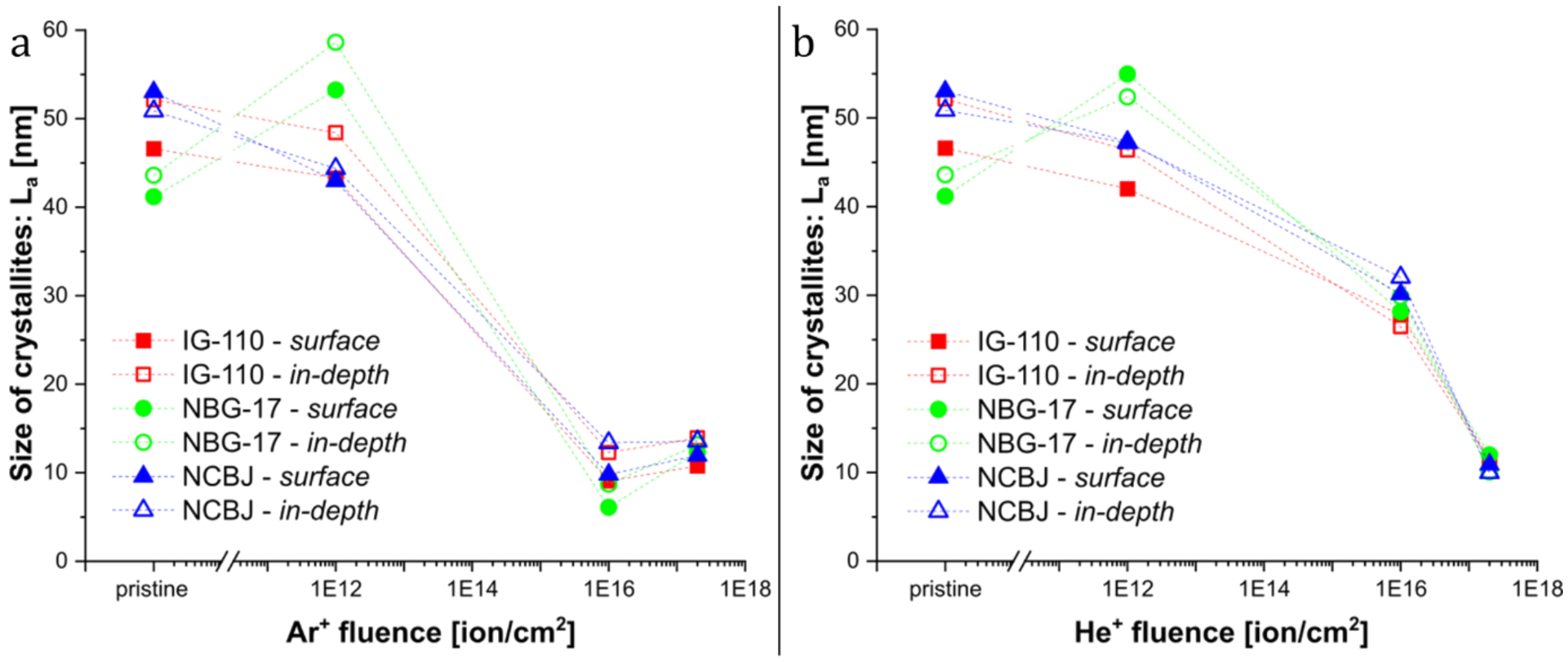

*Figure 8. Estimated size of crystallites calculated for average spectra; dashed lines are provided for better eye guidance*

It needs to be mentioned that results of both in-depth and surface analysis are consistened. No additional effects were were observed dispate higher laser power and different configuration of the in-depth measurements.

## Summary

All analyzed nuclear-grade graphites showed inhomogeneous structure both on the surface and in-depth. This inhomogeneity was also visible in the irradiated samples. In all materials with KMC analysis, it was possible to distinguish regions of significantly different degrees of structural order. In some irradiated samples, parts of a substantially higher degree of order or crystallinity than the rest of the sample were found. Calculated $I_D/I_G$ values, as the indicator of the structural order/disorder, showed graphites to be structurally stable up to specific irradiation fluence. The ratio remains relatively constant from the 0 fluence (pristine sample) up to 1E12 ion/cm$^2$. For further irradiation (1E16 and 2E17 ion/cm$^2$), the value rises considerably up to a certain level. It might be stated that at a threshold value, the material underwent abrupt rather than gradual degradation, visible also by the D and G bands' broadening and vanishing of 2$^{nd}$ order bands (above 2000 cm$^{-1}$). Nevertheless, the $I_D/I_G$ ratio for higher fluences is lower for samples irradiated with lighter $He^+$ ions than for heavier $Ar^+$ since they induce more minor damage in the material. However, the size of crystallites, another indicator of the damage in the material, depicted more differences between the influence of particular ions. Despite all graphite types behaving analogously, the crucial discrepancy is in the progression of reduction of the $L_a$ parameter in the case of $Ar^+$ ions: minimal value is reached at 1E16 ion/cm$^2$ fluence. Meanwhile, for $He^+$, a significant increase in disorder is observed at the highest fluence. Notwithstanding, the performed analysis did not exhibit significant differences between different types of graphites, and the primary type of present defects were vacancies and edge dislocations.

*Acknowledgements*

The authors acknowledge support from the European Regional Development Fund via the Foundation for Polish Science International Research Agenda PLUS program grant No. MAB PLUS/2018/8.

This work is one portion of the studies in the strategic Polish program of scientific research and development work "Social and economic development of Poland in the conditions of globalizing markets GOSPOSTRATEG" part of "Preparation of legal, organizational and technical instruments for the HTR implementation" financed by the National Centre for Research and Development (NCBiR) in Poland (No. Gospostrateg1/385872/22/NCBR/2019).

**Bibliography:**

[1] H. Raza, Graphene Nanoelectronics Metrology, Synthesis, Properties and Applications,, Springer, 2012.

[2] L. Kurpaska, „Structural and mechanical properties of different types of graphite used in nuclear applications," *Journal of Molecular Structure,* tom 1217, p. 128370, 2020.

[3] R. Krishna, „Residual stress measurements in polycrystalline graphite with micro-Raman spectroscopy," *Radiation Physics and Chemistry,* tom 111, pp. 14-23, 2015.

[4] T. Dieing, O. Hollricher and J. Toporski, Confocal Raman Microscopy, Springer, 2010.

[5] G. Giridhar, R. Manepalli and G. Apparao, "Chapter 7 - Confocal Raman Spectroscopy," in *Spectroscopic Methods for Nanomaterials Characterization*, Elsevier, 2017, pp. 141-161.

[6] N. J. Everall, „Confocal Raman microscopy: common errors and artefacts," *Analyst,* tom 135, nr 10, pp. 2512-2522, 2010.

[7] R. Zhang, „In-situ high-precision surface topographic and Raman mapping by divided-aperture differential confocal Raman microscopy," *Applied Surface Science,* tom 546, p. 149061, 2021.

[8] H. Wu, R. Gakhar, A. Chen, S. Lam, C. P. Marshall and R. O.Scarlat, "Comparative analysis of microstructure and reactive sites for nuclear graphite IG-110 and graphite matrix A3," *Journal of Nuclear Materials,* vol. 528, p. 151802, 2020.

[9] E. C. T. Ba, M. R. Dumont, P. S. Martins, B. d. S. Pinheiro, M. P. M. d. Cruz and J. W. Barbosa, "Deconvolution process approach in Raman spectra of DLC coating to determine the sp3 hybridization content using the ID/IG ratio in relation to the quantification determined by X-ray photoelectron spectroscopy," *Diamond and Related Materials,* vol. 122, p. 108818, 2022.

[10] G. Zheng, P. Xu, K. Sridharan and T. Allen, "Characterization of structural defects in nuclear graphite IG-110 and NBG-18," *Journal of Nuclear Materials,* vol. 446, no. 1-3, pp. 193-199, 2014.

[11] A. Eckmann, „Probing the Nature of Defects in Graphene by Raman Spectroscopy," *Nano Letters,* tom 12, nr 8, p. 3925–3930, 2012.

[12] K. Niwase and T. Tanabe, "Defect Structure and Amorphization of Graphite Irradiated by D+ and He+," *Materials Transactions, JIM,* vol. 34, no. 11, pp. 1111-1121, 1993.

[13] B. S. Elman, M. S. Dresselhaus, G. Dresselhaus, E. W. Maby and H. Mazurek, "Raman scattering from ion-implanted graphite," *Physical Review B,* vol. 24, no. 2, pp. 1027-1034, 1981.

[14] Q. Huang, Q. Lei, Q. Deng, H. Tang, Y. Wang, J. Li, H. Huang, L. Yan, G. Lei and R. Xie, "Raman spectra and modulus measurement on the cross section of proton-irradiated graphite," *Nuclear Instruments and Methods in Physics Research B,* vol. 412, p. 221–226, 2017.

[15] T. Nishihara, Excellent Feature of Japanese HTGR Technologies, Japan Atomic Energy Agency: JAEA-Technology, 2018.

[16] J. Ziegler, J. Biersack and U. Littmark, The Stopping and Range of Ions in Solids, New York: Pergamon, http://www.srim.org/, 1985.

[17] M. Hedegaard, C. Matthäus, S. Hassing, C. Krafft, M. Diem and J. Popp, "Spectral unmixing and clustering algorithms for assessment of single cells by Raman microscopic imaging," *Theoretical Chemistry Accounts,* vol. 130, p. 1249–1260, 2011.

[18] J. Ribeiro-Soares, "Structural analysis of polycrystalline graphene systems by Raman spectroscopy," *Carbon,* vol. 95, pp. 646-652, 2015.

[19] N. Dutta, „Self-organized nanostructure formation on the graphite surface induced by helium ion irradiation," *Physics Letters A,* tom 382, nr 24, pp. 1601-1608, 2018.